\documentclass[10pt, final]{IEEEtran}

\usepackage{lipsum}

\usepackage{graphicx}% Include figure files
\usepackage{dcolumn}% Align table columns on decimal point
\usepackage{bm}% bold math. 

\usepackage[english]{babel}
\usepackage{amssymb}
\usepackage{amsmath}
\usepackage{bbm}
\usepackage{nicefrac}
\usepackage{braket}
\usepackage{latexsym}
\usepackage{soul}
\usepackage{braket}
\usepackage{tikz}
\tikzset{
    main/.style={
        draw,circle
        },
    dots/.style={
        },
}
\usepackage{cancel}
\usepackage{caption}
\usepackage{subcaption}
\usepackage{physics}

\usepackage{cite}

\usepackage{hyperref}
\hypersetup{
 colorlinks=true,
 linkcolor=blue,
 filecolor=magenta, 
 urlcolor=cyan,
}
 
\begin{document}

\title{QUBO Resolution of the Job Reassignment Problem}

\author{
Iñigo Perez Delgado{$^1$}, Beatriz García Markaida{$^2$},
Alejandro Mata Ali{$^3$}, Aitor Moreno Fdez. de Leceta{$^4$}

\medskip

{$^1$}i3b Ibermatica, Parque Tecnológico de Bizkaia, Ibaizabal Bidea, Edif. 501-A, 48160 Derio, Spain \\
Email: i.perez.delgado@ibermatica.com

{$^2$}i3b Ibermatica, Parque Tecnológico de Bizkaia, Ibaizabal Bidea, Edif. 501-A, 48160 Derio, Spain \\
Email: b.garcia@ibermatica.com

{$^3$}i3B Ibermatica, Quantum Development Department, Paseo Mikeletegi 5,  20009 Donostia, Spain\\
Email: a.mata@ibermatica.com

{$^4$}i3b Ibermatica, Unidad de Inteligencia Artificial, Avenida de los Huetos,
Edificio Azucarera, 01010 Vitoria, Spain\\
Email: a.moreno@ibermatica.com
}

\maketitle
\begin{abstract}
We present a subproblemation scheme for heuristical solving of the JSP (Job Reassignment Problem). The cost function of the JSP is described via a QUBO hamiltonian to allow implementation in both gate-based and annealing quantum computers. For a job pool of $K$ jobs, $\mathcal{O}(K^2)$ binary variables -qubits- are needed to solve the full problem, for a runtime of $\mathcal{O}(2^{K^2})$. With the presented heuristics, the average variable number of each of the $D$ subproblems to solve is $\mathcal{O}(K^2/2D)$, and the expected total runtime $\mathcal{O}(D2^{K^2/2D})$, achieving an exponential speedup. 
\end{abstract}


\section{\label{sec:intro}Introduction}
QUBO (\textit{Quadratic Unconstrained Binary Optimization}) problems \cite{glover2019tutorial}, which can be solved both by quantum annealing devices and gate-based quantum computers, have been already used to give answer to relevant NP-hard problems \cite{garey1979computers} such as the Travelling Salesperson Problem \cite{villarrodriguez2022TSP,Bonomi2022TSP} or the Hamiltonian Cycle problem \cite{N_lein_2022HamilCycle}. In this paper we will treat one of these QUBOs named the Job Reassignment Problem (JRP). In tha JRP a number $J$ of agents - which can take the form of workers, machines, vehicles- have to be reallocated to a new configuration of jobs due to some unexpected circumstance such as a shift on the production priority or the incapacitation of some of the originally assigned agents. Due to those unexpected causes, there exist a number $I$ of relevant jobs with high priority score which do not have a paired agent. The resolution of the JRP implies finding the agents which better suit those jobs, moving them from their current low-priority jobs to those unassigned high-priority ones. This is done by taking into account some $\mathcal{S}_{ij}$ scores defined between each vacant job $i$ and each agent $j$. The goal of the optimization problem will be to choose the $i,j$ pairs such that the sum of their $\mathcal{S}_{ij}$ is maximized, with the full problem using  $N=J\times I$ binary variables.

At first glance the number of variables required to solve the complete JRP grows linearly with each of two independent parameters $J$ and $I$. However, if typically a fraction $p$ of the total number $K=J+I$ of jobs are emptied, meaning that $I=pK$ and $J=(1-p)K$, both the number $J$ of agents ready to perform the jobs and the number $I$ of unexpectedly empty jobs would escalate equally. Then, $N=J\times I=\mathcal{O}(K^2)$. The search space of possible solutions, and thus the time required to find its optimum, has size $2^{J\times I}=\mathcal{O}(2^{K^2})$. %assuming that the probability of a job being empty is a fixed constant for each problem,

In this paper we present a heuristical approach to the JRP, with several variable-reduction methods that aim to paliate the $N=\mathcal{O}(K^2)$ scaling of the full problem. This scheme divides the full $\mathcal{O}(K^2)$-variable problem into $D$ different $\mathcal{O}(K^2/2D)$-variable subproblems, reducing the search space of each subproblem to size $\mathcal{O}(2^{K^2/2D})$ and thus the total runtime of the $D$ subproblems to $\mathcal{O}(D2^{K^2/2D})$. 

It is important to note that these proportionality figures are expected values, and the true scaling will vary between instances of the problem. Moreover, these methods, being of heuristical nature,  do not guarantee an acceptable solution for all cases of the problem. This does not mean, as it is the case of other heuristics, that no advantage is to be expected from their usage. In fact, the subproblemation allows for agents to be reallocated to originally non-vacant jobs, which is not the case of the original full problem. This expanded answer space suggests that, in some cases, the optimum of the approximated problem can be of better quality than the optimum of the original full hamiltonian. 

The presented methods have been tested in the real context of a joint project between i3b (\textit{Instituto de Innovación Ibermática}) and the ONCE (\textit{Organización Nacional de Ciegos Españoles}).

\section{\label{sec:description}Problem description}

The $\mathcal{S}_{ij}$ coefficients of the cost function of the problem that relate each agent with each vacant job are given by the combination of several values. The first term to take into account is the priority gain $\Delta^\mathcal{P}_{ij}\equiv \mathcal{P}^V_i-\mathcal{P}^C_j$ between the priority of the vacant job $\mathcal{P}^V_i\in(0,1]$ and the priority of the job the agent is currently covering $\mathcal{P}^C_j\in(0,1]$. These priority values are known for each job, being given by the statement of the problem. However, it is allowed for these values to be changed between instances of the resolution of the problem, since a certain job can have different priorities in different contexts. In any case, none of the considered jobs should have a priority of $0$, because in that case the job could just not be considered part of the problem. Priority can have a discrete range of values when each job is ranked in one of $D$ priority categories. In that case, all jobs of the same category will have the same $\mathcal{P}^V_i$ or $\mathcal{P}^C_j$  value, usually $\in\{1,2,...,D\}$.% If the agent is not currently covering any job, $\mathcal{P}^C_j=0$ and $\Delta^\mathcal{P}_{ij}=\mathcal{P}^V_i$.

The second term of $\mathcal{S}_{ij}$ is the affinity gain $\Delta^\mathcal{A}_{ij}\equiv\mathcal{A}^V_{ij}-\mathcal{A}^C_{jj}$ between the personal affinity of agent $j$ with the vacant job $i$, $\mathcal{A}^V_{ij}\in[0,1)$, and the personal affinity of that agent with the job they are currently covering $\mathcal{A}^C_{jj}\in[0,1)$. These affinities are noted collectively as $\mathcal{A}_{kj}$, where $k$ encompasses all $K=I+J$ jobs, vacant or assigned.

The total score is then calculated as
\begin{equation}
\mathcal{S}_{ij}=c^\mathcal{P}\Delta^\mathcal{P}_{ij}+c^\mathcal{A}\Delta^\mathcal{A}_{ij}
\end{equation}
where $c^\mathcal{P}$ and $c^\mathcal{A}$ are the two positive constants that give the relative weight of the optimization terms. They are given by the statement of the problem.

A simple way to model the $\mathcal{A}_{kj}$ personal affinities is by counting the number of times the agent $j$ was assigned to a particular job $k$ in a historical record and mapping that count to a monotonically ascending function such as  

\begin{equation}
\mathcal{A}_{kj}=1-\frac{1}{1+M_{kj}}\;,
\end{equation}

where $M_{kj}$ is the number of times agent $j$ has been assigned to job $k$. Note that, since an assigned agent has been assigned to their current job at least for the current instance, in this model $\mathcal{A}^C_{jj}\in[0.5, 1)$.

%\textcolor{red}{For very large historic records where $M_{kj}$ can grow too much, a reducing proportionality factor could be introduced multiplying $M_{ij}$ and $M_{jj}$. We have not needed it for our implementation, which has used a historic record of about $600$ instances where the maximum values of $M^V_{ij}$ and $M^C_{j}$ were around $400$.   Note that results of more elaborate ways of calculating the $\mathcal{A}_{kj}$ such as classical or quantum machine learning continuous classifiers are also easily implemented in the model.}

\section{\label{sec:integer}Variable selection}

In order to solve this optimization problem one binary variable $x_{ij}\in\{0,1\}$ will be assigned to each `vacant job - agent' pair. If $x_{ij}=1$, then agent $j$ will be reassigned to the vacant job $i$, leaving vacant their current assigned job. If $x_{ij}=0$, then agent $j$ will not be reassigned to job $i$, but could be reassigned for other vacant job. This means that

\begin{equation}
    \text{if } \sum_j x_{ij}=0 \text{ then vacant job $i$ remains vacant,}
\end{equation}

and 

\begin{equation}
    \text{if } \sum_i x_{ij}=0 \text{ then agent $j$ is not reassigned.}
\end{equation}

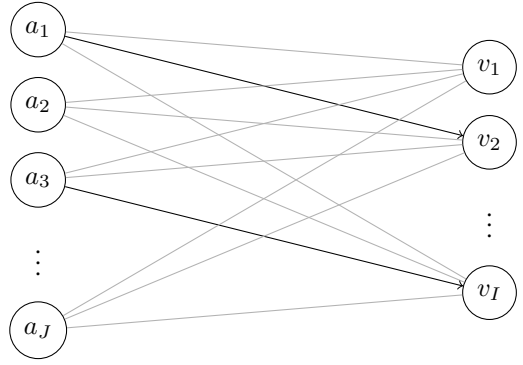
\begin{figure}
\centering
\begin{tikzpicture}[] 
\node[main] (a0) at (0,0) {$a_1$};
\node[main] (a1) [below of=a0] {$a_2$};
\node[main] (a2) [below of=a1] {$a_3$};
\node[dots] (adots) [below of=a2] {$\vdots$}; 
\node[main] (aJ) [below of=adots] {$a_J$};

\node[main] (v0) at (6,-0.5) {$v_1$};
\node[main] (v1) [below of=v0] {$v_2$};
\node[dots] (vdots) [below of=v1] {$\vdots$}; 
\node[main] (vI) [below of=vdots] {$v_I$};

\draw[black!30] (a0) -- (v0);
\draw[black!100, ->](a0) -- (v1);
\draw[black!30] (a0) -- (vI);
\draw[black!30] (a1) -- (v0);
\draw[black!30] (a1) -- (v1);
\draw[black!30] (a1) -- (vI);
\draw[black!30] (a2) -- (v0);
\draw[black!30] (a2) -- (v1);
\draw[black!100, ->] (a2) -- (vI);
\draw[black!30] (aJ) -- (v0);
\draw[black!30] (aJ) -- (v1);
\draw[black!30] (aJ) -- (vI);

\end{tikzpicture} 

\caption{Representation of the full hamiltonian for a problem with $J$ agents -that is, $J$ jobs with assigned agents- and $I$ vacant jobs, where the solution to the problem has involved moving agent $a_1$ and $a_3$ from their original jobs to the originally vacant jobs $v_2$ and $v_I$. Notice how not all vacants are necessarily filled and not all agents are necesarily reallocated. \label{fig:complete}}
\end{figure}

For $J$ agents with assigned jobs and $I$ vacant jobs, the number of binary variables needed to solve the full problem is $N=J\times I$. In Fig. \ref{fig:complete} each of the $x_{ij}\in\{0,1\}$ variables of the full problem is represented with a grey line: if the agent from job $a_j$ has been reallocated to the vacant job $v_i$, then $x_{ij}=1$ and the line is colored black. Else, $x_{ij}=0$ and the line remains grey.

\section{\label{sec:hamiltonian}Hamiltonian construction}

The cost function hamiltonian $H$ will be divided in two parts: the core hamiltonian $H^0$, where we will encode the optimization problem, and the restriction hamiltonian $H^R$ whose terms will effectively reduce the search space to only physically plausible states. Then, $H\equiv H^0+H^R$.

The core hamiltonian takes into account the total score of all active variables, and is composed of two terms, as shown in (\ref{eq:H0}): the priority gain and the affinity gain.

\begin{equation}
H^0=-\sum_{ij}\mathcal{S}_{ij}x_{ij}=-c^\mathcal{P}\sum_{ij}\Delta^\mathcal{P}_{ij}x_{ij}-c^\mathcal{A}\sum_{ij}\Delta^\mathcal{A}_{ij}x_{ij}\;.
    \label{eq:H0}
\end{equation}
Note how even though JRP is a maximization problem, the value of $H^0$ is smaller for high $\Delta^\mathcal{P}_{ij}$ and $\Delta^\mathcal{A}_{ij}$. This happens because the annealing systems decay towards the state of minimum energy of their hamiltonian, so the original maximization cost function has to be translated into its analogous minimization cost function by the introduction of a $-1$ factor. 

The restriction hamiltonian has two terms too, as shown in (\ref{eq:HR}):

\begin{equation}
H^R=H^R_1+H^R_2\;,
    \label{eq:HR}
\end{equation}
where the first term 
\begin{equation}
H^R_1=\lambda_1^R\sum_{i}\left(\sum_{j}x_{ij}-0.5\right)^2
    \label{eq:HR1}
\end{equation}
ensures, for a large enough $\lambda_1^R>0$, that each job $i$ can be done by at most one agent and the second term
\begin{equation}
H^R_2=\lambda_2^R\sum_{j}\left(\sum_{i}x_{ij}-0.5\right)^2
    \label{eq:HR2}
\end{equation}
makes, for a large enough $\lambda_2^R>0$, each agent $j$ to be reassigned to at most one job. Allowing the restriction coefficient to be the fractionary number $0.5$ allows the number of active binary variables of the sums inside the parentheses to be either $0$ or $1$, since those are the integer values that are closer to $0.5$ \cite{Perez2022}, without introducing any dummy variables. Using only integer restriction coefficients would force the introduction of $I+J$ dummy variables.

The complete hamiltonian is then, merging (\ref{eq:H0}), (\ref{eq:HR1}) and (\ref{eq:HR2}),
\begin{equation}
\begin{split}
    H = -\sum_{ij}&\bigg[c^\mathcal{P}\left(\mathcal{P}^V_i-\mathcal{P}^C_j\right)+c^\mathcal{A}\left(\mathcal{A}^V_{ij}-\mathcal{A}^C_{jj}\right)\bigg]x_{ij}\\
    &+\lambda_1^R\sum_{i}\bigg(\sum_{j}x_{ij}-0.5\bigg)^2\\
    &+\lambda_2^R\sum_{j}\bigg(\sum_{i}x_{ij}-0.5\bigg)^2\;,
    \label{eq:Htot}
\end{split}
\end{equation}
summed over the $i,j$ pairs represented by the lines of the graph.

\section{\label{sec:reduction}Heuristical variable reduction and Problem segmentation}

In order to diminish the needed number of qubits $N$, which in the full problem of Fig. \ref{fig:complete} equals $J\times I$, certain simplifications will be taken.

Firstly, only changes with $\mathcal{S}_{ij}>0$ will be taken into account. Changes with a negative score will be ignored, even if they would allow a second, positive-scored change which resulted in a net gain. It is assumed that the initial distribution is already in a sensible state, and as such it would be difficult to obtain a gain with that kind of second-order movement. 

Secondly, changes with $\Delta^\mathcal{P}_{ij}<0$ will also be ignored, since the ultimate goal of the optimization is to maximize the total priority of the assigned jobs. 

Assuming about half of the $i,j$ pairs have $\mathcal{S}_{ij}>0$ or $\Delta^\mathcal{P}_{ij}<0$ -since the two quantities are strongly correlated, otherwise $\nicefrac{3}{4}$ of the pairs would be assumed-, these simplifications reduce the expected number of qubits needed to approximately $N\approx (J\times I) /2$, thus reducing the search space to a size of $\sqrt{2^{J\times I}}$ and achieving a quadratical speedup.

Moreover, ignoring changes with $\Delta^\mathcal{P}_{ij}<0$ allows us to divide the problem into smaller subproblems. These subproblems will be generated in function of the $\mathcal{P}^V_i$ priorities of the vacant values. If those values are discrete, being able to take $D$ different values, then for each value one subproblem will be created. If the values are continuous, they can be grouped in $D$ intervals of length $1/D$. Then for the $d$th subproblem, only those vacant jobs with $\mathcal{P}^V_i\geq 1-\frac{d}{D}$ are considered. After having solved the $d=1$ subproblem, those vacants which have been succesfully reassigned will not be included in the $d=2$ subproblem. However, those jobs that have been emptied as a result of their agent being reassigned will be included as vacant jobs. This means all $\mathcal{A}_{kj}$ are potentially needed for the resolution, not only the $\mathcal{A}^V_{ij}$ and $\mathcal{A}^C_{jj}$ of the original full problem. Moreover, as $\Delta^\mathcal{P}_{ij}<0$ variables are ignored, only those agents with jobs with $\mathcal{P}^C_j\leq{\mathcal{P}^V_d}_{MAX}$ will be considered, where ${\mathcal{P}^V_d}_{MAX}$ is the maximum priority between all considered $\mathcal{P}^V_i\geq 1-\frac{d}{D}$. 

\begin{figure}
\centering
\begin{tikzpicture}[] 
\node[main] (a0) at (0,0) {$a_1$};
\node[main] (a1) [below of=a0] {$a_2$};
\node[main] (a2) [below of=a1] {$a_3$};
\node[main] (a3) [below of=a2] {$a_4$};

\node[main] (v0) at (6,-0.5) {$v_1$};
\node[main] (v1) [below of=v0] {$v_2$};
\node[main] (v2) [below of=v1] {$v_3$};

\draw[black!100, ->](a0) -- (v0);
\draw[black!30] (a0) -- (v1);
\draw[black!30] (a0) -- (v2);
\draw[black!30] (a1) -- (v0);
\draw[black!30] (a1) -- (v2);
\draw[black!100, ->] (a2) -- (v2);
\draw[black!30] (a3) -- (v2);

\node[dots] (darrow12) at (3,-3.5) {$\big\Downarrow$};

\node[main] (a4) at (0,-4.5) {$a_2$};
\node[main] (a5) [below of=a4] {$a_4$};

\node[main] (v3) at (6,-4) {$v_4$};
\node[main] (v4) [below of=v3] {$v_5$};
\node[main] (va2) [below of=v4] {$a_3$};

\draw[black!30] (a4) -- (va2);
\draw[black!100, ->](a4) -- (v3);
\draw[black!30] (a4) -- (v4);
\draw[black!100, ->] (a5) -- (va2);

\end{tikzpicture} 

\caption{Representation of the subproblem hamiltonians for a problem with $J=4$ agents and $I=5$ vacant jobs, divided into $D=2$ subproblems. The first solves for the high $\mathcal{P}^V_i$ vacants $\{v_i\}_1=\{v_1,v_2,v_3\}$, and then a second subproblem takes care of the remaining $\{v_4,v_5\}$, as well as the newly-emptied $\{a_3\}$, which makes $\{v_i\}_2=\{v_4,v_5,a_3\}$. Meanwhile $\{a_1\}$, a specially low-priority job, is not even considered for this playoff.  After resolution, the list of jobs with assigned agents has become $\{v_1, v_4, v_3, a_3\}$.
\label{fig:subproblems}}
\end{figure}
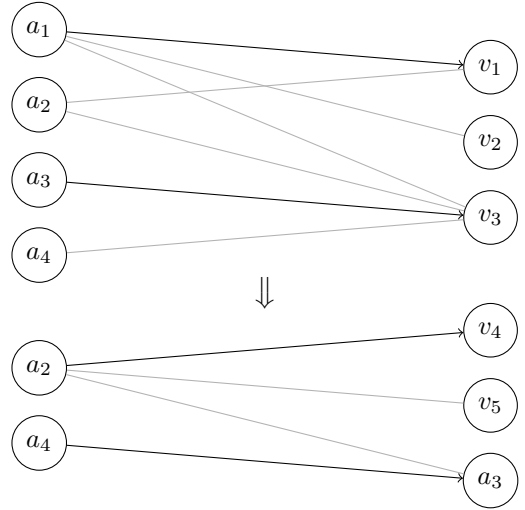

We can also estimate the effect of the subproblemation on the size of the search space and time. It segmentates the $I$ vacant jobs of the complete problem into $D$ subproblems, giving an average value of $\langle I_D\rangle=I/D$ empty jobs for subproblem. As one can see in Fig. \ref{fig:subproblems}, The number of available agents is also reduced with each step, with each step having $\alpha\langle I_D\rangle$ less agents than the previous one, where $\alpha$ is the average proportion of vacant jobs that are covered on each subproblem. Then, the expected number of agents of the $d$th subproblem, for $d\in \{1,...,D\}$, would be $J-\alpha D(d-1)$, for an average of $\langle J_D\rangle=J-\alpha D(D-1)/2$ over the $D$ subproblems. In total, each subproblem will have an expected size of $\langle N_D\rangle=\langle J_D\rangle\times\langle I_D\rangle=JI/D -\alpha I(D-1)/2=\mathcal{O}(K^2/D).$ This in turn makes the search space of each subproblem size $\mathcal{O}(2^{K^2/D})$, and the total runtime of the $D$ subproblems $\mathcal{O}(D2^{K^2/D})$. 

With all the heuristics combined, the search space of each subproblem is expected to be reduced to size $\mathcal{O}(2^{K^2/2D})$, and the total runtime of the $D$ subproblems to $\mathcal{O}(D2^{K^2/2D})$.

In Fig. \ref{fig:subproblems} a representative example of such a subproblemation secheme is shown, for a $D=2$ case. Notice how, in the example, $a_j$ jobs emptied in the first subproblem can be taken by other agents in the second, lower-priority subproblem. In this case, subproblemation has reduced a size $J\times I=4\times5=20$ problem down to a $4\times3=12$ and a $2\times3=6$ problem: not only the total size is smaller $(18<20)$, but each of the subproblems can be executed in a machine with a significantly lower number of logical qubits. Moreover, applying our first variable-reduction criterion, only $x_{ij}$ variables whose $\mathcal{S}_{ij}$ score are positive are taken into account. This is represented by the omission of some of the gray connections on the figure, In this example, this reduces the needed number of qubits of the two subproblems to $7$ and $4$ respectively.
\section{\label{sec:algoritm}Summary of the quantum algorithm}

\textbf{Inputs:} 
\begin{itemize}
    \item The $\{a_j\}$ list of the $J$ jobs with assigned agents -including the identity of the assigned agent $j$.
    \item The $\mathcal{P}^C_j$ priorities of the $\{a_j\}$ jobs,
    \item The $\{v_i\}$ list of the $I$ vacant jobs.
    \item The $\mathcal{P}^V_i$ priorities of the $\{v_i\}$ jobs,
    \item The $\mathcal{A}_{ik}$ affinities between each agent $j$ and each job $k$. These affinity scores will be used in the form of the affinity of agent $j$ with the vacant job $i$, $\mathcal{A}^V_{ij}$ and the affinity of of agent $j$ with the job they are covering $\mathcal{A}^C_{jj}$.
    \item The $c^\mathcal{P}$ and $c^\mathcal{A}$ coefficients regulating the relative weights of the priority and affinity terms.
\end{itemize}

\textbf{Procedure:}
\begin{enumerate}
    %\item Make a copy of $\{a_j\}$ called $\{a_j\}^0$ in order to store the initial job assigned to each agent.
    \item Decide, depending on the caracteristics of $\{v_i\}$ and $\mathcal{P}^V_i$, the number $D$ of subproblems: 
    \begin{enumerate}
        \item If the values of $\mathcal{P}^V_i$ are discrete, then for each value one subproblem will be created. 
        \item If the values of $\mathcal{P}^V_i$ are continuous, the elements of $\{v_i\}$ can be grouped into $D$ intervals of average length $1/D$.
    \end{enumerate}
    \item Divide $\{v_i\}$ into the $D$ different sublists $\{v_i\}_d$, with $d$ ordered from high to low $\mathcal{P}^V_i$. Each $\{v_i\}_d$ will contain vacant jobs with priorities $\mathcal{P}^V_i\in[{\mathcal{P}^V_d}_{min},{\mathcal{P}^V_d}_{MAX}]$.
    \item For $\delta\in\{1,...,D\}$, repeat iteratively steps a) $\rightarrow$ e):
    \begin{enumerate}
         \item  Knowing only $\Delta^\mathcal{P}_{ij}>0$ changes can happen, create a sublist $\{a_j\}_\delta$ with only those assigned jobs with $\mathcal{P}^C_j<{\mathcal{P}^V_\delta}_{MAX}$.  
         \item Create a graph connecting all elements of $\{a_j\}_\delta$ with all elements of $\{v_i\}_\delta$, and then remove all $i,j$ connections with $\mathcal{S}_{ij}\leq 0$. 
         \item Find the minimum of the $H$ hamiltonian described by the graph, which has the QUBO form given by (\ref{eq:Htot}),  taking into account that each connection is a representation of one $x_{ij}$ binary variable.
         \item Update the $\{a_j\}$ list of jobs which have an assigned worker, taking into account the new locations of the reallocated workers by removing new vacants and adding newly filled jobs. Note that the elements of the list change and thus the $j$ indices refer to different jobs in different iterations. However, the number of elements $J$ does not change.
         \item Update all the $\{v_i\}_d$ to include the jobs the reallocated agents have just left vacant. Only $\{v_i\}_d$ with $d \in\{\delta+1,...D\}$ will need an update.
         \begin{enumerate}
            \item For discrete values of $\mathcal{P}^V_i$, each of the new vacants will naturally be included in the $\{v_i\}_d$ group that corresponds to their $\mathcal{P}^C_j$.
            \item For continuous values of $\mathcal{P}^V_i$ sometimes jobs with $\mathcal{P}^C_j\in[{\mathcal{P}^V_\delta}_{min},{\mathcal{P}^V_\delta}_{MAX}]$ will be left vacant. In that case, they should be included into the inmediately subsequent list of $\{v_i\}_{\delta+1}$, for them to be included in the subproblem of the next iteration.  ${\mathcal{P}^V_{\delta+1}}_{MAX}$ will then need to be updated, taking the value of the highest $\mathcal{P}^C_j$ included.
        \end{enumerate}
    \end{enumerate}   
\end{enumerate}

\textbf{Output:} An updated version of the $\{a_j\}$ list with the jobs that have ended up with an assigned agent -and the identity of that agent for each job. 

\section{\label{sec:conclusions}Conclusions}

In this paper we described the heuristical subproblemation scheme developed for solving the Job Reassignment Problem posed in the joint project between i3b (\textit{Instituto de Innovación Ibermática}) and the ONCE (\textit{Organización Nacional de Ciegos Españoles}). 

This algorithm, as all heuristical methods, does not guarantee a speedup or a solution of quality for all possible instances of the problem. However, the performance of the method for an average problem can be calculated, proving an exponential speedup over the resolution of the full problem without subproblemation. Moreover, the subproblemation scheme uniquely allows for agents to be reallocated to originally non-vacant jobs, which means that in some cases the given answer can be of better quality than the optimum of the original full problem.

The presented algorithm works over a QUBO hamiltonian cost function to allow resolution by both gate-based and annealing quantum computers, as well as classical and quantum-inspired resolution methods. As it divides the QUBO hamiltonian of the full problem into several smaller problems which are still in QUBO form, the advantage of the subproblemation scheme is device-agnostic and stacks with the advantages of other QUBO resolution methods. As usual, all the advantages of the method remain when using the Ising forms of the hamiltonians.

\section*{\label{sec:aknowledgements}Acknowledgments}
We thank ONCE INNOVA for the project proposal which kickstarted this work.

The research leading to this paper has received funding from
the Q4Real project (Quantum Computing for Real Industries),
HAZITEK 2022, no. ZE-2022/00033.

© 2023 IEEE. Personal use of this material is permitted. Permission from IEEE must be obtained for all other uses, in any current or future media, including reprinting/republishing this material for advertising or promotional purposes, creating new collective works, for resale or redistribution to servers or lists, or reuse of any copyrighted component of this work in other works.

%\section*{\label{sec:interests}Competing interests}
%The authors declare no competing interests. 

\bibliographystyle{IEEEtran}
\bibliography{apssamp}

\end{document}